# Performing Path Integral Molecular Dynamics Using Artificial Intelligence Enhanced Molecular Simulation Framework

Cheng Fan[1,2], Maodong Li[2], Sihao Yuan[1,2], Zhaoxin Xie[1,2], Dechin Chen[2], Yi Isaac Yang[2*], Yi Qin Gao[1,2,3*]

1. Institute of Theoretical and Computational Chemistry, College of Chemistry and Molecular Engineering, Peking University, Beijing 100871, China
2. Institute of Systems and Physical Biology, Shenzhen Bay Laboratory, Shenzhen 518107, China
3. New Cornerstone Science Laboratory, College of Chemistry and Molecular Engineering, Peking University, Beijing, China
*Corresponding authors: Yi Isaac Yang (yangyi@szbl.ac.cn), Yi Qin Gao (gaoyq@pku.edu.cn)

**Abstract:**
This study employed an artificial intelligence-enhanced molecular simulation framework to enable efficient Path Integral Molecular Dynamics (PIMD) simulations. Owing to its modular architecture and high-throughput capabilities, the framework effectively mitigates the computational complexity and resource-intensive limitations associated with conventional PIMD approaches. By integrating machine learning force fields (MLFFs) into the framework, we rigorously tested its performance through two representative cases: a small-molecule reaction system (double proton transfer in formic acid dimer) and a bulk-phase transition system (water-ice phase transformation). Computational results demonstrate that the proposed framework achieves accelerated PIMD simulations while preserving quantum mechanical accuracy. These findings show that nuclear quantum effects can be captured for complex molecular systems, using relatively low computational cost.

## 1 Introduction

Nuclear quantum effects (NQEs) play a vital role in determining both static equilibrium properties and dynamic processes across chemical[1,2] and materials systems[3,4], particularly in systems involving light nuclei such as hydrogen[5]. For instance, classical molecular dynamics (MD) simulations systematically overestimate free energy barriers in proton transfer processes[6] and fail to capture kinetic isotope effects. The inclusion of NQEs yields significantly improved agreement with experimental measurements.[7] This quantum mechanical nature of nuclei is indispensable for elucidating many phenomena ranging from enzymatic catalytic mechanisms[8] to the relative stability of polymorphic crystal phases[9], especially when light atoms such as hydrogen are involved. However, establishing the precise characterization of NQEs is still a pivotal challenge in contemporary theoretical and computational chemistry.[10]

The path integral molecular dynamics (PIMD) framework provides a rigorous and efficient approach for modeling nuclear quantum behavior through isomorphic mapping of quantum particles onto fictitious classical ring polymers in an extended phase space.[11–13] This formalism enables systematic sampling of quantum-statistical ensembles, thereby capturing essential quantum mechanical features including zero-point energy, quantum fluctuations, and tunneling.[11] On the other hand, the fidelity of PIMD simulations critically depends on the accuracy of underlying potential energy surfaces (PESs). While *ab initio* methods such as density functional theory (DFT) can achieve high accuracy in PES construction, their integration with PIMD incurs in many cases prohibitive computational costs.[14] For typical studies, it demands computational resources tens to hundreds of times greater than equivalent classical MD simulations.[10] This disparity becomes further exacerbated when contrasting first-principles calculations against empirical force field. Such prohibitive computational demands impose severe constraints on both accessible system sizes and achievable simulation timescales, forcing researchers to make suboptimal compromises between computational accuracy, system sizes, and simulation timescales.

Recent advances in machine learning force fields (MLFFs) have triggered a paradigm shift in molecular simulations by enabling DFT-level accuracy with dramatically reduced computational cost.[15] Contemporary MLFF implementations typically use deep molecular models[16], i.e., molecular conformation encoding models based on deep neural networks, to fit high-dimensional PESs. Deep molecular models can be roughly classified into two categories based on feature representation strategies: descriptor-based models and graph neural network (GNN)-based models. The descriptor-based was initiated by Behler-Parrinello neural network (BPNN), where handcrafted symmetry functions were used to encode atomic environments, which were subsequently processed through neural networks to predict atomic energies.[17] Subsequent development of descriptor-based models exemplified by the Deep Potential model[18] has been widely used in systems such as condensed matter. In parallel, molecular models with GNN architectures emerged from the message-passing neural network established by Gilmer et al.,[19] providing an explicit molecular representation through graphs where nodes and edges correspond to atoms and bonds, respectively. Innovative advancements have been made in resent GNN-based MLFFs. Models such as NequIP[20] and Allegro[21] employs equivariant architectures, which preserve fundamental physical symmetries while achieving state-of-the-art accuracy. MolCT[22], developed by our group, utilizes a multi-head ego-attention mechanism to capture the atomic environment. It was demonstrated to possess high fitting ability in establishing force fields for reactions in MD simulations.[23] Furthermore, the integration with classical force fields enables machine learning/molecular mechanics (ML/MM) multiscale simulations, thereby enhancing the description accuracy of chemical systems to approach realistic environmental conditions.[24,25]

Current mainstream MLFFs, predominantly developed using Python-based AI frameworks (e.g., TensorFlow and PyTorch), often face compatibility challenges when interfacing with conventional MD simulation packages written in compiled languages (C/C++/Fortran). To address this interoperability issue, several established MD packages have implemented specialized application programming interfaces (APIs) for MLFF integration. For instance, LAMMPS[26] provides interface for popular models such as Deep Potential and SchNet[27]. HOOMD-TF has been specifically implemented to facilitate the integration of machine learning methods within the HOOMD-blue framework.[28] While conventional MD software requires substantial code modification for MLFF integration, MD engines implemented based on artificial intelligence (AI) framework permits the seamless integration with MLFFs like JaxMD[29], TorchMD[30]. Similarly, MindSPONGE simulation packages was developed based on an artificial intelligence-enhanced molecular simulation (AIMS) framework.[31] However, MLFF-based PIMD simulations still present unique implementation challenges. Only few software supports the PIMD simulations. The current widely-used implementation for PIMD simulations is the i-PI package developed by Cerrotti et. al..[32] It adopts a Python-based client-server architecture where server process handles nuclear coordinate evolution via equations of motion and client processes (external MD packages) handle the remaining process through inter-process communication. For instance, through integration with the LAMMPS simulation package, one can run PIMD simulations using Deep Potential model.[33] Despite its compatibility with popular MD engines such as LAMMPS and CP2K[34], this decoupled design introduces non-negligible communication latency and exerts a limit on the scaling efficiency.

To address the aforementioned limitations, this study introduces an innovative PIMD simulation protocol through the integration of the AIMS framework using a MLFF built with graph field network (GFN). The Python-centric unified simulation environment fundamentally resolves cross-platform interoperability constraints. The modular design permits direct implementation of advanced techniques, including PIMD and enhanced sampling methods such as metadynamics (MetaD) [35], without requiring external software interfaces. To validate the practical application of this approach, two model systems were investigated: the double proton transfer process in the formic acid dimer and the relative phase stability between cubic ice and liquid water. The results demonstrate the ability of the protocol for performing efficient PIMD simulations.

## 2 Methods
### 2.1 Path Integral Molecular Dynamics

Considering the Hamiltonian of an N-atom system under the Born-Oppenheimer approximation:

$$\hat{H} = \sum_{i=1}^{N} \frac{\hat{p}_i^2}{2m_i} + V(r_1, r_2, \ldots, r_N)$$

with the corresponding partition function given by:

$$Z = \text{Tr}[\exp(-\beta\hat{H})]$$

Applying the Suzuki-Trotter decomposition to the imaginary-time evolution operator yields a reformulated partition function:

$$Z = \lim_{P\to\infty} \int d^{NP}\mathbf{r}\, e^{-\beta \hat{H}_{\text{eff}}}$$

where the effective Hamiltonian $\hat{H}_{\text{eff}}$ is defined as:

$$\hat{H}_{\text{eff}} = \sum_{i=1}^{N}\sum_{k=1}^{P}\left[\frac{p_i^{(k)2}}{2m_i'} + \frac{1}{2}m_i\omega_P^2\left(r_i^{(k+1)} - r_i^{(k)}\right)^2 + \frac{V\left(r_i^{(k)}\right)}{P}\right]$$

where $\omega_P = \sqrt{P}/(\beta\hbar)$. To improve numerical stability in practical implementations, the normal mode or staging transformation is employed to decouple the evolution of individual beads, resulting in a partitioned effective Hamiltonian:

$$\hat{H}_{\text{eff}} = \hat{H}_p + \hat{U}$$

$$\hat{H}_p = \sum_{i=1}^{N}\sum_{k=1}^{P}\left[\frac{p_i^{(k)2}}{2m_i'} + \frac{1}{2}m_i\omega_P^2\left(r_i^{(k+1)} - r_i^{(k)}\right)^2\right]$$

$$\hat{U} = \sum_{i=1}^{N}\sum_{k=1}^{P}\frac{V\left(r_i^{(k)}\right)}{P}$$

For a thermodynamic property $O(r_1, r_2, \ldots, r_N)$, its estimator with a finite number of beads is:

$$\langle O \rangle = \frac{1}{Z_P}\text{Tr}\left[\hat{O}e^{-\beta\hat{H}_{\text{eff}}}\right] = \frac{\int d^{NP}r\, \hat{O}(r_1, r_2, \ldots, r_N)e^{-\beta\hat{H}_{\text{eff}}}}{\int d^{NP}r\, e^{-\beta\hat{H}_{\text{eff}}}}$$

These expressions highlight the critical role of the bead number P in resolving nuclear quantum effects. While increasing P improves the accuracy of the quantum mechanical representation, it concomitantly escalates computational demands. This inherent trade-off between quantum fidelity and computational cost distinguishes path-integral simulations from classical molecular dynamics approaches. Due to the high computational cost, it is challenging to run PIMD simulations in large systems at *ab initio* level.

**2.2 Graph Field Network for PIMD simulation**

Typical GNN-based models treat the molecular systems as graphs where atoms serve as nodes and inter-atomic relationships serve as edges. Employing atomic coordinates and atomic types as typical fundamental input features, each node aggregates information from its neighboring nodes through iterative message-passing mechanisms, subsequently updating its atomic representation. Finally, all node and edge representations are readout to predict the properties of the system. Thus, the GNN-based force field uses this model to predict the potential energy of the system, while the interatomic forces computed via automatic differentiation of the energy with respect to atomic coordinates. This force fields based on automatic differentiation offers advantages such as transferability and simplified force field construction, but it also introduces some critical limitations. A substantial computational overhead will be incurred when calculating atomic force by back-propagation. In addition, the fitting of the total potential energy of the system limits parallel computing of the model.

Here, we propose an improved approach for GNN-based molecular modeling using graph field networks (GFN), which can significantly improve the computational speed of GNN-based force fields. GFN is a translationally invariant and rotationally equivariant network architecture, originally developed for predicting particles displacement in diffusion-based generative models.[36] Within our MLFF framework, GFN is re-purposed to directly predict the forces acting on atoms within molecules. We consider forces to be an "edge" property, thus can be decoded directly from edge representations. A couple of operations showed below are designed to ensure messages are well-transferred between nodes and edges.

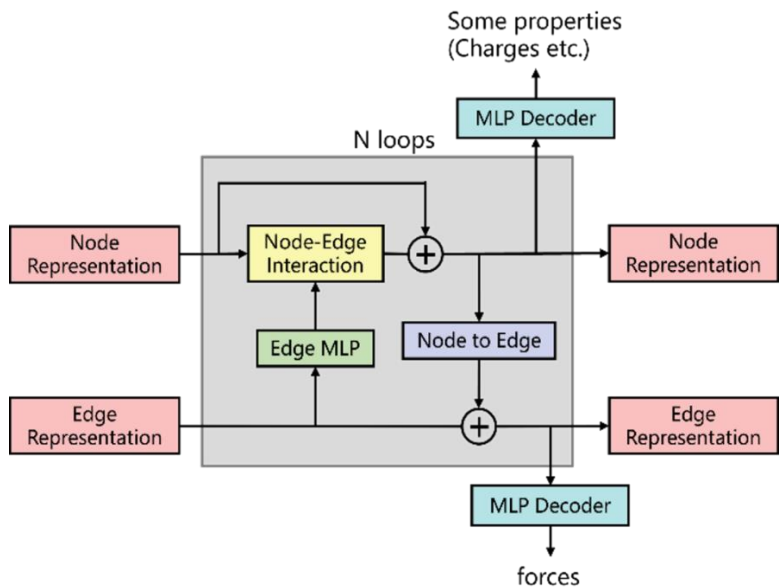

Fig 1. The architecture of Graph Field Network.

$$m_i^{(l)} = \phi_m\left(e_{ij}^{(l)}\right)$$

$$n_i^{(l+1)} = n_i^{(l)} + \phi_v\left(n_i^{(l)}, m_i^{(l)}\right)$$

$$e_{ij}^{(l+1)} = e_{ij}^{(l)} + \text{agg}\left(n_i^{(l)}, n_j^{(l)}\right)$$

Edge information obtained from a MLP module is integrated into node representations through the interacting module. This procedure is followed by an MLP module, which passes messages from nodes to edges. The force readout originates from the updated edge representation:

$$F_{ij} = c_{ij} * d_{ij} * \phi_f\left(e_{ij}^{(l)}\right)$$

Eliminating the need for gradient backpropagation in simulations substantially enhances processing speed and thus expands its applicability to larger-scale systems. This approach circumvents the recursive derivative calculations typically required in energy-predict methods, thereby reducing computational complexity and memory overhead. Here, we build on the MolCT model by using GFN as the readout representation of the model to directly fit the atomic forces of the simulation system.

## 2.3 AI-enhanced Molecular Simulation Framework

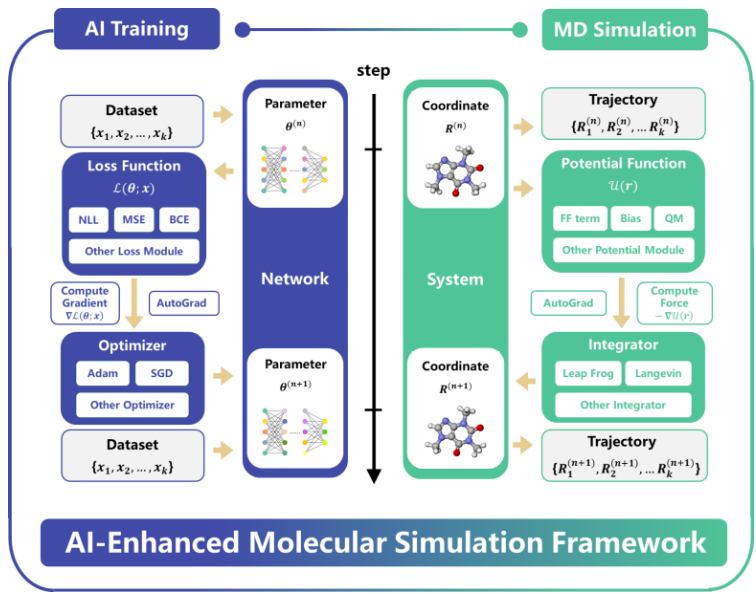

Fig 2. Illustration of AI training and MD simulation. Their analogy results in the AI-enhanced molecular simulation frameworks.

Here, we use a program framework based on AIMS to run PIMD simulations. As shown in Fig.2, this MD simulation framework has a similar workflow to AI training, whereby the simulation "system" corresponds to neural "network", "potential function" mirrors "loss function", and "integrator" aligns with "optimizer" algorithms. This systematic mapping drove the reimplementation of conventional MD simulations within the AI-inspired architecture. The framework features: modular architecture design, automated parallelization, and native support for automatic differentiation, which makes it naturally suitable for performing PIMD simulations on the MLFF-based potential energy function.

The modular architecture of the program enables potential energy functions to be implemented as interchangeable modules with standardized interfaces, allowing flexible substitution and mixed usage between classical force fields, first-principles methods (such as DFT through the PySCF[37,38] interface), and MLFFs. Automated parallelization enables high-throughput execution of concurrent trajectory simulations, establishing the foundation for PIMD simulations with independent mode evolution. Notably, by using GPUs, multiple data is processed with the same instruction in parallel, significantly reducing the execution time. Furthermore, native automatic differentiation capability enables efficient computation of gradients from harmonic potentials, providing essential infrastructure for primary PIMD implementations.

Within our AI-architected simulation framework, molecular systems (including coordinates $R$ with potential energy $U(R)$ are encapsulated as modular components named "WithEnergyCell". These components are subsequently integrated with "Sponge" to form the core computation engine. For first-principles MD (FPMD) implementations, quantum mechanical computations are integrated via the "WithForceCell" module, which directly interfaces with PySCF, giving the energy and atomic forces through first-principles electronic structure evaluations as output. Chart 1 shows a typical simulation code.

```
from sponge import PySCFForceCell
from sponge import WithForceCell
from sponge import RunOneStepCell
from sponge import Sponge

# Wrap energy and forces from PySCF with simulation system
pyscf_frc = PySCFForceCell(system, pyscf_option, mol_parameters, ks_parameters)
sys_with_frc = WithForceCell(system, pyscf_frc)

# Set simulation engine
onestep = RunOneStepCell(None, sys_with_frc , optimizer.integrator)
md = Sponge(onestep)

#Start running
md.run(1000)
```

Chart 1. Code Snippet 1: Running FPMD in the AIMS framework

PIMD capabilities are implemented through two different modules based on the computational requirements. For primitive PIMD, inter-bead harmonic coupling potentials are incorporated as pluggable energy terms within "WithEnergyCell", with forces automatically computed through the framework's built-in auto-differentiation engine. To perform normal-mode/staging transformations, one only needs a substitution of the temporal integrator module while maintaining identical potential interfaces. Chart 2 shows a typical simulation code:

```
from sponge import WithEnergyCell
from sponge import NMPIMD
from sponge import Sponge

# Wrap potential energy with simulation system
sys_with_ene = WithEnergyCell(system, potential)

# Set simulation engine
Integrator = NMPIMD(system, time_step)
md = Sponge(sys_with_ene, Integrator)

#Start running
md.run(1000)
```

Chart 2. Code Snippet 2: Running PIMD in the AIMS framework

This architecture mitigates the need for redundant code modifications – new methodologies necessitate only targeted module substitutions while maintaining existing workflow components. The programming paradigm fundamentally shifts developers' focus from low-level implementation logistics to high-level algorithm design, and thus effectively decouples novel method development from framework-specific integration challenges. The software codes can be downloaded from the Gitee repository: https://gitee.com/helloyesterday/mindsponge/tree/develop

## 2.5 Metadynamics

Owing to its modular design, the framework can also integrate the functionality of enhanced sampling methods to accelerate the sampling of rare times in PIMD simulations. For example, we implement the well-tempered metadynamics (WT-MetaD)[39] which accelerates rare event exploration through collective variables (CVs). The CV $s(R)$ is defined as a low-dimensional projection of the system coordinates $R$. Within the WT-MetaD framework, a time-dependent bias potential $V_G(s,t)$ is introduced to the Hamiltonian, constructed as a superposition of Gaussian kernels deposited along the CV trajectory. Crucially, the Gaussian height self-adapts according to the accumulated bias potential:

$$V_G(s,t) = \int_0^t \omega_0 \exp\left(-\frac{\beta V_G(s,t')}{\gamma - 1}\right) \exp\left(-\frac{|s - s(t')|^2}{2\sigma}\right) dt'$$

where $\omega_0$ denotes the initial Gaussian height, $\sigma$ controls the kernel width, and $\gamma$ serves as the bias factor. [40]This adaptive protocol ensures systematic convergence of the bias potential:

$$V(s) \propto -\left(1 - \frac{1}{\gamma}\right) F(s)$$

with

$$F(s) = -\frac{1}{\beta} \ln p_0(s) = -\frac{1}{\beta} \ln \langle \delta(S(r) - s) \rangle$$

For quantum systems, the free energy surface (FES) along CV $S(r_1, r_2, \ldots, r_N)$ is evaluated via path-integral reweighting:

$$F_{quantum}(s) = -\frac{1}{\beta} \ln \left[\frac{1}{P} \sum_{k=1}^{P} \langle \delta(S(r^{(k)}) - s) \rangle\right] = -\frac{1}{\beta} \ln \left[\frac{1}{P} \sum_k^P e^{-\beta F^{(k)}(s)}\right]$$

Through the above equation quantum nuclear effects are captured as ensemble statistics of individual path-integral beads.

## 3 Results and Discussion

### 3.1 Double Proton Transfer in Formic Acid Dimer

First, we tested out protocol on the proton transfer process in a small-molecule system,

formic acid dimer (FAD). The formic acid dimer represents an ideal model system for testing and validating computational methods because it is simple enough to allow for highly accurate reference calculations. The double proton transfer in this system has been a focal point of numerous theoretical studies, as it serves as a prototypical example where NQEs play significant roles. The double proton transfer process in this system can be described by the following CV $s_{dimer}$:

$$s_{dimer} = d_{O^{(1)}H^{(1)}} - d_{O^{(2)}H^{(1)}} + d_{O^{(4)}H^{(2)}} - d_{O^{(3)}H^{(2)}}$$

where $d_{XY}$ represents the interatomic distance between atoms X and Y. Atomic labels are shown in the Figure 3.

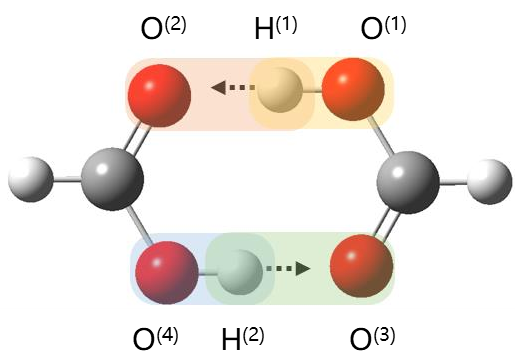

Fig. 3 Molecular representation of a double-proton transfer process in a FAD. Corresponding collective variable is listed below the representation

To establish a rigorous benchmark for classical nuclear situations, we first conducted FPMD simulations for 100 ps at 200 K. The simulations predict an energy barrier being 4.53 kcal/mol along the reaction coordinate for the double proton transfer process. This result serves as the reference point for evaluating the accuracy of our MLFFs in capturing classical nuclear behavior.

Subsequently, we trained two types of MLFFs which are based on MolCT and MolCT-GFN respectively at the same level of theory used for FPMD simulations. All models were trained using a dataset comprising 2918 distinct conformations. As illustrated in Figure 4, the training process for both models exhibited stable convergence. The MolCT model achieved a value of ~1.99 kcal/(mol·Å) for the root-mean-square error (RMSE) for forces, while the MolCT-GFN model attained a remarkably low force RMSE of ~0.48 kcal/(mol·Å). Both models demonstrated excellent agreement between predicted and true forces.

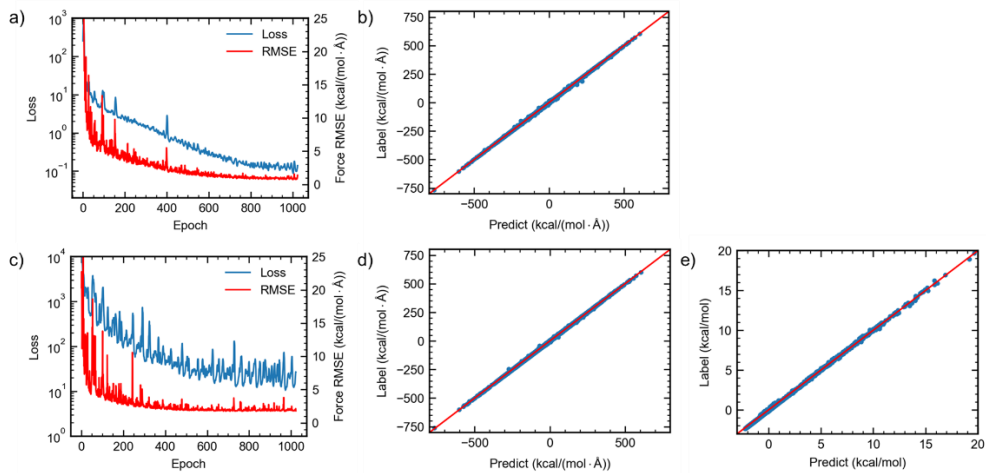

Fig 4. Training curves for the GFN network (top); training curves for the energy-based model (bottom).

To assess the reliability of these force fields for thermodynamic calculations, we performed 150 ps classical MD simulations using both MLFFs to construct the FES of the double proton transfer process. The comparative FES profiles in Figure 5a reveal good agreement between MLFFs and reference quantum mechanical results with the energy barrier all being 4.53 kcal/mol along the reaction coordinate. This numerical correspondence demonstrates that our MLFFs successfully reproduce quantum-level accuracy in classical MD simulations. From a computational efficiency perspective, the MolCT-GFN architecture has a simulation speed of 8 M steps/day, 1.8× faster than standard MolCT model (4.5 M steps/day) and 44.4× more efficient than baseline FPMD (0.18 M steps/day) in MD simulation tasks. The results above collectively highlight the enhanced computational efficiency and maintained accuracy of our MLFF implementations.

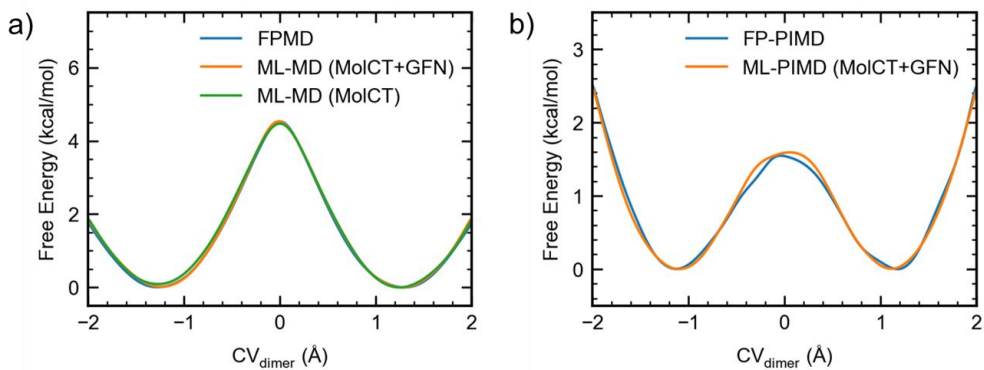

Fig 5. Free energy profile of double proton transfer in the formic acid dimer obtained from a) FPMD and FP-PIMD; b) FPMD, ML-MD (MolCT+GFN), ML-MD (MolCT) ML-PIMD (MolCT+GFN), and FP-PIMD.

Then, we performed the first-principles PIMD (FP-PIMD) to investigate the role of NQEs in the double proton transfer process. With all simulation settings being the same, the FP-PIMD simulation results predict a 67% decrease in activation barrier from 4.53 kcal/mol to 1.52 kcal/mol. This quite dramatic energy barrier reduction demonstrates that NQEs critically facilitate proton transfer by lowering the activation energy and enhancing the probability of quantum tunneling. Our findings are consistent with the seminal work of Aran et al., who reported comparable threefold barrier reductions when incorporating NQEs in FAD proton transfer systems. [41]

To verify the MolCT-GFN force field can accurately reproduce NQEs at a much lower computational cost, we performed machine learning PIMD (ML-PIMD) simulations. A simulation time of 150 ps was employed. Figure 5b shows the potential energy profiles derived from both simulations using our computational framework. The ML-PIMD simulations well reproduced the FP-PIMD results, which yielded a barrier height of 1.59 kcal/mol deviating by only 0.07 kcal/mol from the first-principles reference. This good agreement demonstrates that our MLFF framework successfully captures quantum mechanical effects while maintaining ab initio accuracy. Crucially, the MolCT-GFN model achieves this quantum-level accuracy with little increase in the computational cost. Compared to FP-PIMD (0.006 M steps/day), our ML-PIMD approach offers a 1200× faster simulation speed (7.2 M steps/day). This efficiency makes ML-PIMD a practical tool for studying NQEs in larger and more complex systems where direct first-principles PIMD would be computationally prohibitive.

Table 1 shows a systematic comparison of the differences in simulation efficiency. While FP-PIMD suffered a 97% speed reduction, ML-PIMD maintained 90% computational efficiency. Crucially, when taking the MLFF training into consideration, the total computational cost still remains ~30 of times lower than FP-PIMD requirements for equivalent scientific outcomes. Such a gain in computational efficiency becomes more evident when longer time and large-scale simulations are required.

Table 1. Simulation speed of different methods

| Method | Speed | Accelerate |
| --- | --- | --- |
| FPMD | 0.18M steps/day | 1.0× |
| ML-MD (Energy) | 4.5M steps/day | 25.0× |
| ML-MD (GFN) | 8M steps/day | 44.4× |
| Method | Speed | Accelerate |
| ML-FPMD | 0.006M steps/day | 1.0× |
| ML-PIMD (GFN) | 7.2M steps/day | 1200.0× |

* All tests were done on a NVIDIA A100 (80 GB) GPU

## 3.2 Water-ice Phase Transition in Bulk Systems

We subsequently used the above protocol to investigate a more complex phase transition system: the ice-water phase transition. The ice-water phase transition is of fundamental importance in condensed matter physics, and has profound implications for global climate patterns and Earth's energy dynamics. A number of studies has been published elucidating the molecular mechanisms driving this transformation.[42,43] However, little work has been done to study NQEs in ice-water phase transition processes as a result of the vast conformational space and extensive temporal scales involved. Such a problem is thus almost computationally prohibitive for conventional FP- PIMD simulations. Here, we use AI-enhanced molecular modeling software to perform PIMD simulations with the MolCT-GFN force field to study the NQEs of this system. In order to achieve a reversible ice-water phase transition process in the simulations, we use the MetaD enhanced sampling method with a CV based on X-ray diffraction (XRD) intensities[44], which is very effective for ice-water phase transitions: [42,43,45]

$$S_{water} = \frac{1}{N} \sum_j^N \sum_i^N f_i(Q) f_j(Q) \frac{\sin(Q \cdot R_{ij})}{Q \cdot R_{ij}}$$

where $Q = 4\pi \sin\theta / \lambda$ is the scattering vector, $f_i(Q)$ and $f_j(Q)$ are the atomic scattering factors, and $R_{ij} = |\vec{R}_i - \vec{R}_j|$ represents the distance between two atoms.

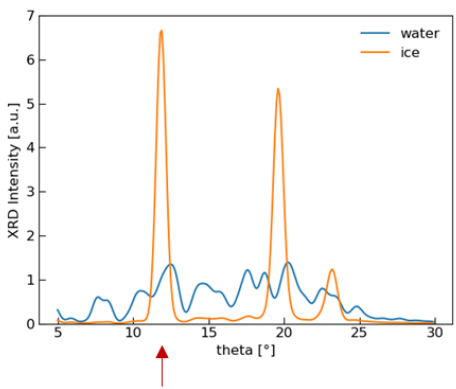

Fig 6. XRD intensity of water and ice at different angles. Corresponding collective variable is listed below the representation

To construct a self-consistent machine learning force field (MLFF), we utilized a comprehensive dataset consisting of 1,204 configurations, where each configuration represents a system containing 64 water molecules arranged in an orthogonal box. Figure 7a displays the evolution of the loss function for the training set and the root mean square error (RMSE) of forces for the validation set over successive training epochs. As the training process progresses, the loss function gradually decreases and stabilizes, finally exhibiting only marginal fluctuations without a notable downward trend, while the RMSE on the validation set also remains steady. Therefore, the training process has achieved convergence without evidence of overfitting. Furthermore, we performed test evaluations on the entire dataset using the trained force field, as shown in Figure 7b. The test results demonstrate that the force field achieves an RMSE of only 2.50 kcal/(mol·Å) in force prediction, highlighting the force field's high predictive accuracy, which is essential for the stringent requirements of high-precision molecular dynamics simulations.

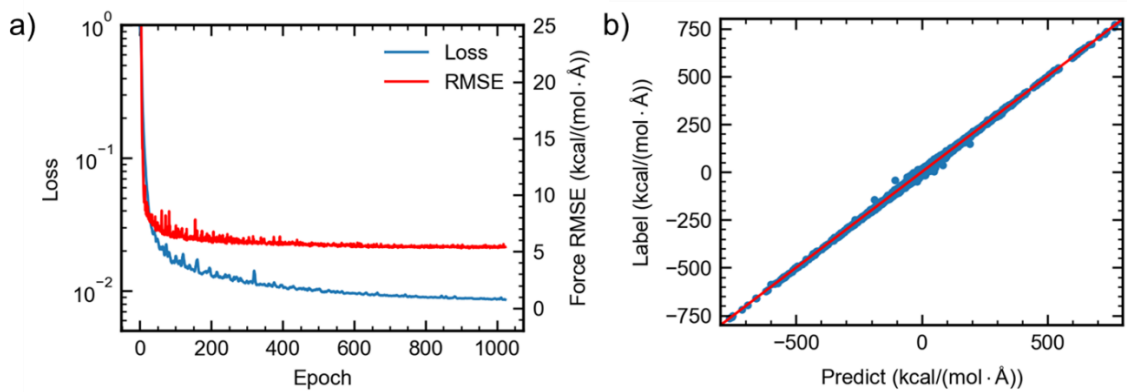

Fig 7. The curves of the loss function on the training set and the RMSE of forces on the validation set against the training epochs (left), and the plot showing the relationship between the predicted and true values of forces across the entire dataset (right).

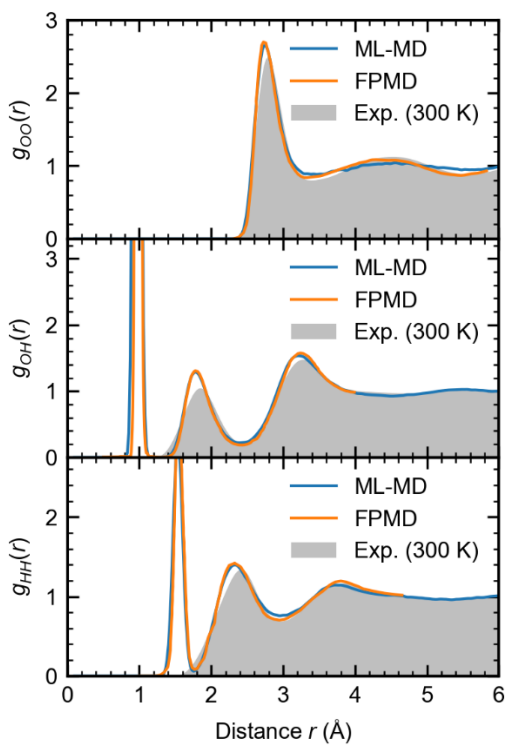

Fig 8. The RDFs of O-O, O-H, and H-H in water obtained from ML-MD, FPMD, and experiments. The FPMD data are sourced from Chen[46] and Zheng[47], and the experimental data are sourced from Soper[48].

To further validate the performance of the constructed MLFF, we performed MD simulations on a system consisting of 64 water molecules under periodic boundary conditions. The radial distribution function (RDF), a key and commonly used structural property that provides critical insights into the molecular arrangements and interactions, was selected as the primary evaluation metric to assess the MLFF's accuracy in a realistic simulation environment. Figure 6 presents a comprehensive comparison of the O–O, O–H, and H–H RDFs derived from our GFN model alongside the FPMD benchmarks. Good agreement is observed between the GFN-based RDF profiles and those obtained from FPMD simulations, with only minor deviations evident in certain peak intensities. This result demonstrates the ability of MLFF to accurately reproduce the core features of DFT calculations. We note here that the first peak of the O–O RDF occurs at 2.73 Å for both GFN and FPMD simulations, which is slightly shifted from the experimental value of 2.79 Å.

To account for NQEs, we employed PIMD simulations. The comparison between classical MD and PIMD in Figure 9 reveals distinct differences in the structural properties induced by NQEs. Specifically, the O-O RDF remains largely unchanged, indicating that nuclear quantum effects do not significantly alter the oxygen-oxygen pair correlations in this system. In contrast, the O-H RDF exhibits notable modifications: its first peak broadens and shifts slightly toward shorter interatomic distances. These observations are consistent with previous studies, where NQEs were found to soften hydrogen bonding interactions and enhance molecular mobility.

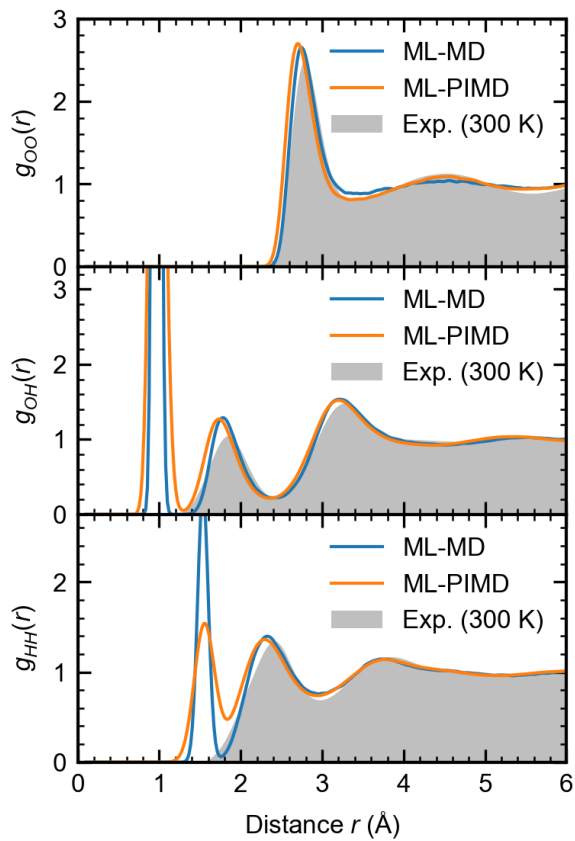

Fig 9. RDFs of O-O, O-H, and H-H in water, obtained from ML-MD and ML-PIMD simulations

To systematically explore the influence of NQEs on the relative stability of ice and water phases, we conducted simulations at different temperatures. At 315 K, for example, the simulation results clearly delineate distinct regions corresponding to liquid water and cubic ice phases, visualized in Figure 10a. Specifically, CV values around 1 were characteristic of liquid water, while those near CV=6 were indicative of cubic ice. Notably, our simulation trajectory exhibited continuous and frequent transitions between these two states. This dynamic behavior highlights the delicate balance between the two phases under the given conditions. To quantitatively assess the thermodynamic stability, we analyzed the free energy profile as a function of CV, as shown in Figure 11b. The free energy difference ($\Delta F$) is defined as $\Delta F = F_{ice} - F_{water}$, calculated as the difference between the lowest free energy minima of the two phases.

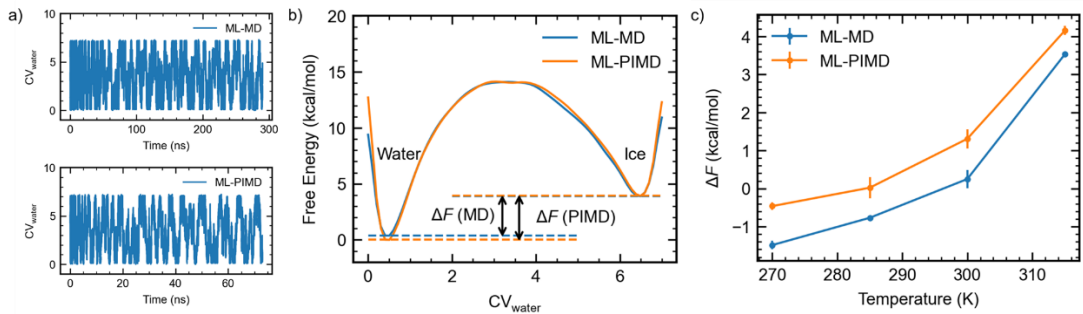

Fig 10. a) CV changes against time at 315 K; b) Free energy surface obtained at 315 K, the definition of $\Delta F$ is labeled in the graph; c) The free energy differences ($\Delta F$) at different temperatures.

We systematically evaluated ΔF across a range of temperatures. At all four temperatures investigated, ML-PIMD simulations yield a higher ΔF compared to classical MD simulations. This observation suggests that NQEs likely contribute to the preferential stabilization of the liquid water phase over the cubic ice phase. The melting point, defined by the condition:

$$\Delta F = F_{ice} - F_{water} = 0$$

was determined from these calculations. Figure 10c on the free energy differences at different temperatures shows that the inclusion of NQEs result in a significant decrease in the melting point (approximately 10 K compared to the classical system. The calculated melting points for the quantum and classical systems are ~285 K and ~295 K, respectively. This significant shift underscores the critical role of nuclear quantum effects in modulating the phase behavior of water under these conditions.

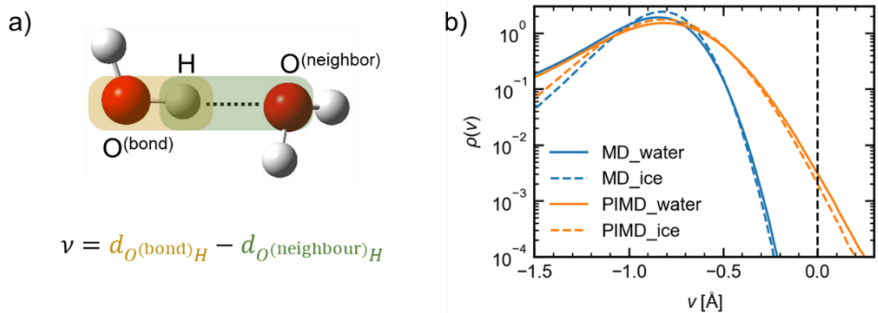

Fig 11. a) The illustration of proton transfer coordinate. Corresponding calculation is listed below. b) Proton transfer coordinate at 315 K analyzed for water and ice phase

respectively from ML-MD and ML-PIMD simulations. The black dashed line divides the region where the hydrogen is closer to the bonding oxygen or the nearest non-bonding oxygen.

We next investigated the differences between the two phases by analyzing the proton transfer coordinate ($v$), a critical metric for evaluating hydrogen bonding characteristics. The proton transfer coordinate is defined as $v = d_{O^{(bond)}H} - d_{O^{(neighbor)}H}$, where $d_{O^{(bond)}H}$ is the distance between a hydrogen atom and its bonding oxygen, and $d_{O^{(neighbor)}H}$ is its distance to the nearest non-bonding oxygen, as shown in Figure 12a. Our comparative analysis of quantum and classical simulations revealed distinct distribution patterns as shown in Figure 12b. Notably, the inclusion of NQEs significantly enhanced the delocalization of hydrogen atoms in water, resulting in a small but notable fraction of configurations where *v > 0*, unlike classical molecular dynamics (MD) simulations, which exhibited no such configurations (*v < 0* for all cases). This observation provides direct evidence that NQEs lead to stronger and more robust hydrogen bonding networks in the quantum system compared to the classical treatment.

Furthermore, a phase-dependent discrepancy was observed: water exhibits a higher probability of hydrogen delocalization compared to ice, and this difference becomes even more pronounced when NQEs are accounted for. This observation suggests that quantum effects play a more significant role in modulating hydrogen bond strength in liquid water than in its crystalline ice counterpart. Importantly, these trends were consistently observed across all investigated temperatures, indicating a fundamental relationship between quantum effects, hydrogen bonding, and phase stability. These findings provide new insights into the role of NQEs in explaining the physicochemical behaviors of water and ice, such as the isotope-dependent melting point variations (e.g., $\Delta T_m = 3.8\ K$ between $H_2O$ and $D_2O$).

## 4 Conclusion

We here present a systematic investigation of the ML-PIMD simulation within AIMS framework. Our methodological advancement centers on the development of a new

MLFF termed GFN. This approach was designed to bypass the gradient calculation thus improving the simulation efficiency. Through the integration of PIMD algorithms within the AIMS framework, we established a unified computational platform that enables full quantum dynamical simulations. Validation studies on two prototypical systems demonstrate the method's dual capability in accuracy and efficiency. The results of FAD system shows that GFN well reproduced first-principles level results and displayed great accelerations over *ab initio* methods as well as the energy-based methods. In the test of bulk $H_2O$ systems, we conducted the phase transition simulations at different temperatures and successfully discovered the NQEs resulting in a decrease of about 10 K in the melting points.

The GFN-based force field provides an effective tool for performing PIMD simulations at low cost. In one hand, the GFN framework represents a new type of force fields. By eliminating the need for automatic differentiation to compute forces, GFN achieves significant computational speedups while maintaining first-principles-level accuracy. This advancement potentially enables simulations of mesoscale biomolecular assemblies previously inaccessible. Although the current implementation focuses on force predictions, which are sufficient for most MD applications, future work will incorporate energy prediction that could be vital in certain cases through distillation techniques, ensuring the framework's applicability to a broader range of problems.

In the other hand, the AIMS framework introduced here is a robust platform for investigating NQEs in complex molecular systems. A unique advantage of this framework is its ability to perform multi-trajectory simulations with minimal computational overhead compared to single-trajectory calculations. This efficiency arises from its highly parallelizable architecture, which naturally aligns with the requirements of PIMD simulations. As a result, the framework is particularly well-suited for studying quantum tunneling phenomena in enzymatic catalysis, mass-dependent isotope effects in crystalline materials, and other processes where NQEs play a critical role.

## 5 Acknowledgement

The authors thank Yanheng Li, Yifan Li, Yijie Xia, and Qiming Sun for useful discussion. Computational resources were supported by Shenzhen Bay Laboratory supercomputing centre and CPL high performance computing. This research was supported by the National Science and Technology Major Project (No. 2022ZD0115003), the National Natural Science Foundation of China (22273061 to Y.I.Y., and T2495221, 92353304, 21927901 to Y.Q.G.) and New Corner-stone Science Foundation (to Y.Q.G.).

# Supplementary Information
## S1 Validation of PIMD implementation

We tested our implementation of PIMD algorithms first on some model questions. The one-dimensional quantum harmonic oscillator has the Hamiltonian:

$$\hat{H} = \frac{\hat{p}^2}{2m} + \frac{1}{2}m\omega^2\hat{x}^2$$

It has the exact resolution of ensemble-average energy:

$$\langle E \rangle_{quantum} = \frac{1}{2}\hbar\omega \coth\left(\frac{\hbar\omega}{2k_BT}\right)$$

and the energy for classical treatment is:

$$\langle E \rangle_{classical} = k_BT$$

For P-bead system, the quantum energy is calculated as:

$$\langle E_P \rangle_{quantum} = 3k_BT \sum_{l=1}^{P} \frac{1}{1 + \left[\left(\frac{2k_BT}{\hbar\omega}\right)P\sin\left(\frac{l\pi}{P}\right)\right]}$$

Our results on the bead-num test, the temperature test and the time-step test show the correct implementation:

T=10 K, $\omega$=12.44, $\Delta t$=0.5 fs

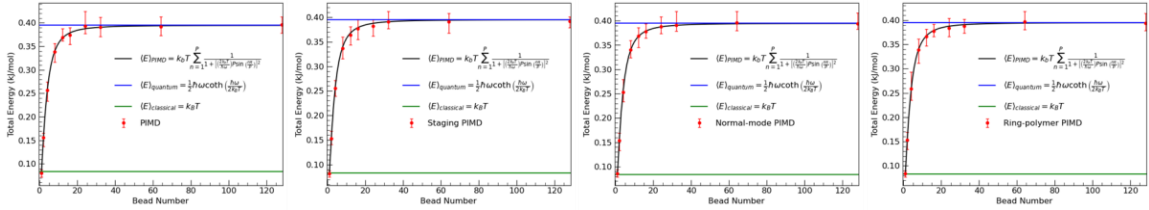

P=128, $\omega$=12.44, $\Delta t$=0.5 fs

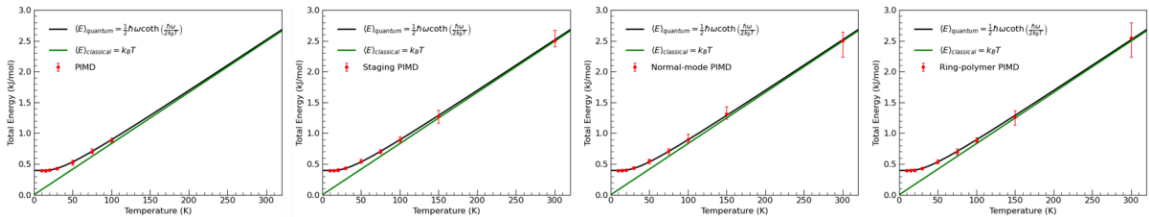

T=10 K, P=128, $\omega$=12.44

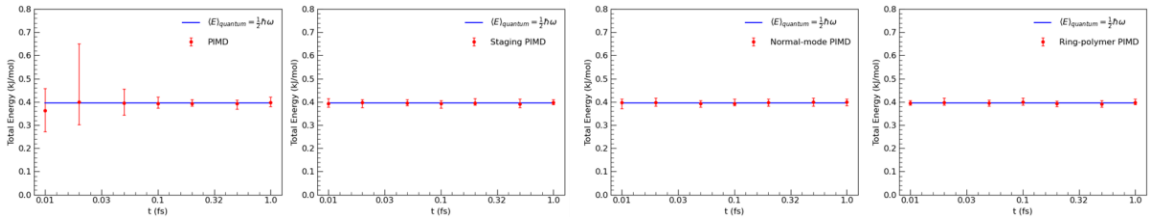

We further compared our PIMD simulation results for single water molecule with that obtained from i-PI, showing excellent consistence:

T=300 K, $\Delta t$=0.5 fs, P=32

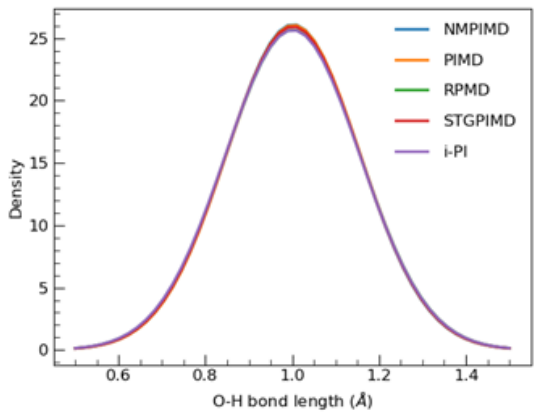

**S2 Computational Details for FAD**

The training of the force field adhered to the procedure previously established by Yuan et al. for the construction of reactive force fields. \cite{yuan2024} We opted for the PBE0 functional in conjunction with the 6-31G** basis set. Quantum mechanical (QM) calculations were executed utilizing the PySCF software package. Notably, we only collected forces from QM results as labels for ML training process. The hyperparameters of MLFF are listed in Table S1.

Table S1. Hyperparameters of MLFF

| System | Feat. Dim. | Act. Func. | Cutoff | N_inter | N_heads | Readout_iter |
|---|---|---|---|---|---|---|
| MolCT+GFN | 128 | SiLU | 0.6 nm | 3 | 8 | 3 |
| MolCT (Energy) | 128 | SiLU | 0.6 nm | 3 | 8 | 3 |

| Param Num | N_epoch | Batch_size | Optimizer | Loss Func. | Learning rate | Metrics |
|---|---|---|---|---|---|---|
| 388609 | 1024 | 32 | Adam | MSE | Exponential Decay | RMSE |
| 266753 | 1024 | 32 | Adam | MSE | Exponential Decay | RMSE |

All simulations were performed on the NVIDIA A100 (80 GB) GPU using AI-styled simulation framework.

**Sampling** To rapidly and accurately construct the free energy profile, we employed Metadynamics as the enhanced sampling method. For the formic acid dimer, the distance differences were selected as the collective variable:

$$CV_{dimer} = d_{O^{(1)}H^{(1)}} - d_{O^{(2)}H^{(1)}} + d_{O^{(4)}H^{(2)}} - d_{O^{(3)}H^{(2)}}$$

The Metadynamics simulations were performed with bias added every 100 steps, using an initial height of 1.0 kJ/mol and a bias factor of 10.

Simulation details are listed below.

| Simulation | Ensemble | Temperature | Time step | Total time |
|---|---|---|---|---|
| ML-MD | | | | 0.15 ns |
| FPMD | NVT | 200 K | 0.5 fs | 0.1 ns |
| ML-PIMD | | | | 0.15 ns |
| FP-PIMD | | | | 0.1 ns |

| Thermostat | Time constant |
|---|---|
| Langevin | 0.1 ps |
| PILE-L | |

### S3 Computational Details for Water

The training of the force field adhered to the procedure previously established by Yuan et al. for the construction of reactive force fields. \cite{yuan2024} We employed the SCAN functional along with norm-conserving pseudopotentials. All calculations were carried out by Quantum Espresso\cite{qe2020}. Notably, we only collected forces from QM results as labels for ML training process. The hyperparameters of MLFF are listed in Table S2.

Table S2. Hyperparameters of MLFF

| System | Feat. Dim. | Act. Func. | Cutoff | N_inter | N_heads | Readout_iter |
|---|---|---|---|---|---|---|
| Water | 64 | SiLU | 0.6 nm | 3 | 8 | 3 |

| Param Num | N_epoch | Batch_size | Optimizer | Loss Func. | Learning rate | Metrics |
|---|---|---|---|---|---|---|
| 100097 | 1024 | 32 | Adam | MSE | Exponential Decay | RMSE |

All simulations were performed on the NVIDIA A100 (80 GB) GPU using AI-styled simulation framework.

**Sampling** Metadynamics was employed as the enhanced sampling method. We choose XRD3D as the collective variable for enhanced sampling:

$$CV_{water} = \frac{1}{N} \sum_{j}^{N} \sum_{i}^{N} f_i(Q) f_j(Q) \frac{\sin(Q \cdot R_{ij})}{Q \cdot R_{ij}}$$

where $Q = 4\pi \sin\theta / \lambda$ is the scattering vector, $f_i(Q)$ and $f_j(Q)$ are the atomic scattering factors, and $R_{ij} = |\vec{R}_i - \vec{R}_j|$ represents the distance between two atoms. The $\theta$ and $\lambda$ were set as $11.95°$ and 1.54 Å. The Metadynamics simulations were performed with bias added every 0.05 ps, using an initial height of 1.0 kJ/mol and a bias factor of 60.

Simulation set ups are listed in the table.

| Simulation | Task | System | Density | Ensemble |
|---|---|---|---|---|
| ML-MD | RDF | 64 $H_2O$ | Water phase | NVT |

| | Phase Transition | Ice phase |
| --- | --- | --- |
| ML-PIMD | RDF | Water phase |
| | Phase Transition | Ice phase |

| Temperature | Time step | Total time | Thermostat | Time constant |
| --- | --- | --- | --- | --- |
| 330 K | | 0.25 ns | Langevin | |
| 270, 285, 300, 315 K | 0.5 fs | 250 ns | | 0.1 ps |
| 330 K | | 0.25 ns | PILE-L | |
| 270, 285, 300, 315 K | | 75 ns | | |

Given that the inclusion of NQEs introduces additional computational complexity due to the need for multiple beads to represent quantum nuclei, determining the appropriate number of beads is essential. Our convergence analysis revealed that at least 32 beads are required to achieve stable and reliable results. While increasing the bead count enhances precision, it simultaneously escalates computational costs, as illustrated in Figure 9. After carefully balancing accuracy and computational efficiency, we selected 32 beads for our final PIMD simulations.

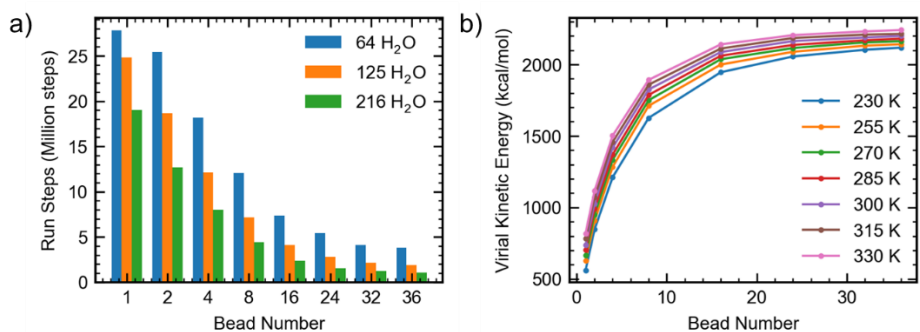

Fig. a) Simulation steps per day when using different numbers of beads in PIMD simulation. All speed tests were performed on the NVIDIA A100 (80 GB); b) Virial estimator of kinetic energy when using different numbers of beads in PIMD simulation.